\documentclass{aa}
\usepackage{txfonts}
\usepackage{graphicx}

\begin{document}

\title{Luminosity and mass functions of galactic open clusters: II. NGC~4852
\thanks{Based on observations carried out at the European Southern
Observatory, La Silla, Chile}.}

\author{G. Carraro\inst{1,2,3}, G. Baume\inst{1,4}, 
 G. Piotto\inst{1}, R. A. M\'endez\inst{2,5}, and
 L. Schmidtobreik\inst{5}}

\offprints{G. Carraro}

\mail{gcarraro@das.uchile.cl}

\institute{
Dipartimento di Astronomia, Universit\`a di Padova, Vicolo
dell'Osservatorio 5, I-35122 Padova, Italy
\and
Departamento de Astronom\'ia, Universidad de Chile, Casilla 36-D,
Santiago, Chile
\and 
Astronomy Department, Yale University, P.O. Box 208101, New Haven,
CT 06520-8101 USA
\and 
Facultad de Ciencias Astron\'omicas y Geof\'{\i}sicas de la UNLP,
             IALP-CONICET, Paseo del Bosque s/n, La Plata, Argentina
\and 
ESO-Chile, Casilla 19001, Santiago, Chile
}

\date{Received xxx / Accepted xxx}

\abstract{We present wide field deep $UBVI$ photometry for the 
previously unstudied open
cluster NGC~4852 down to a
limiting magnitude $I\sim24$, obtained from
observations taken with the Wide Field Imager camera on-board
the MPG/ESO 2.2m telescope at La Silla (ESO, Chile).
These data are used to obtain the first estimate of the cluster basic parameters,
to study the cluster
spatial extension by means of star counts, and to derive the Luminosity (LF)
and Mass Function (MF).  The cluster radius turns out to be $5.0\pm1.0$ arcmin. 
The cluster emerges clearly from the field down to V=20 mag. At fainter magnitudes, it 
is completely confused with the general
Galactic disk field.
The stars inside this region define a young open cluster 
(200 million years old)
1.1 kpc far from the Sun (m-M = 11.60, E(B-V) = 0.45).
The Present Day Mass Functions (PDMF) from
the $V$ photometry is one of the most extended in mass insofar
obtained, and can be
represented as a power-law with a slope $\alpha = 2.3\pm0.3$ and
(the Salpeter (1955) MF in this
notation has a slope $\alpha = 2.35$), in the mass range $3.2 \leq
\frac{m}{m_{\odot}} \leq 0.6$.\\ 
Below this mass, the MF cannot be
considered as representative of the cluster MF,
as the cluster merges with the field and therefore the MF
is the result of the combined effect of
strong irregularities in the stellar background
and
interaction of the cluster with the dense Galactic field.
The cluster total mass at
the limiting magnitude results to be 2570$\pm$210 M$_{\odot}$.
 }

\authorrunning{Carraro et al.}  \titlerunning{NGC 4852}

\maketitle

\begin{figure*}
\centering
\includegraphics[width=14cm]{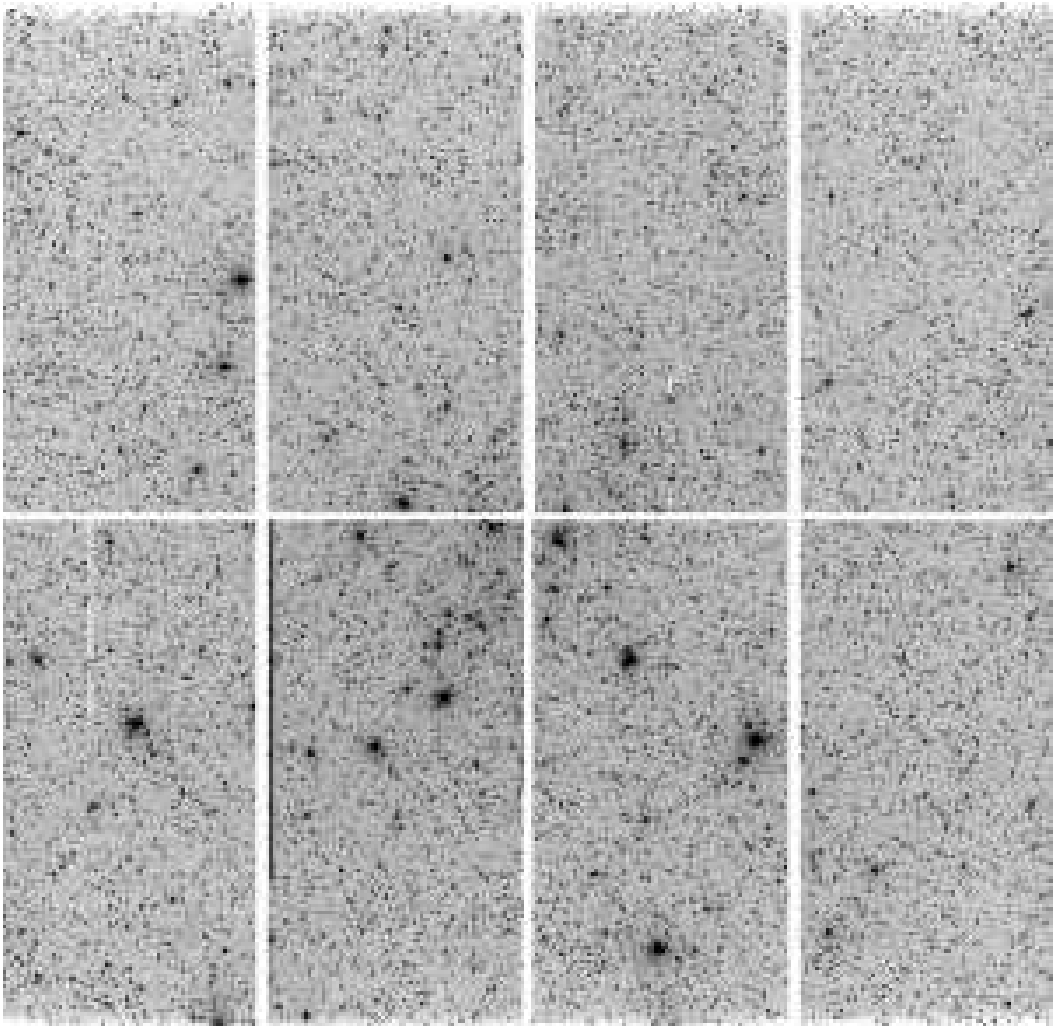}
\caption{A deep WFI image of the region around NGC~4852 taken in V filter.
The field center is RA = 13$^{h}$00$^{m}$06$^{s}$ and 
Dec=-59$^{o}$36$^{\prime}$02$^{\prime\prime}$(J2000). Each chip
covers  an area of $8^{\prime}.12 \times 16^{\prime}.25$, 
while the full field of
view is of $34^{\prime} \times 33^{\prime}$. The cluster center is located
not far from the center of the image, and the cluster has an apparent  radius
of about 2-3 arcmin.}
\label{ccd}
\end{figure*}

\begin{figure}
\centering
\includegraphics[width=9cm]{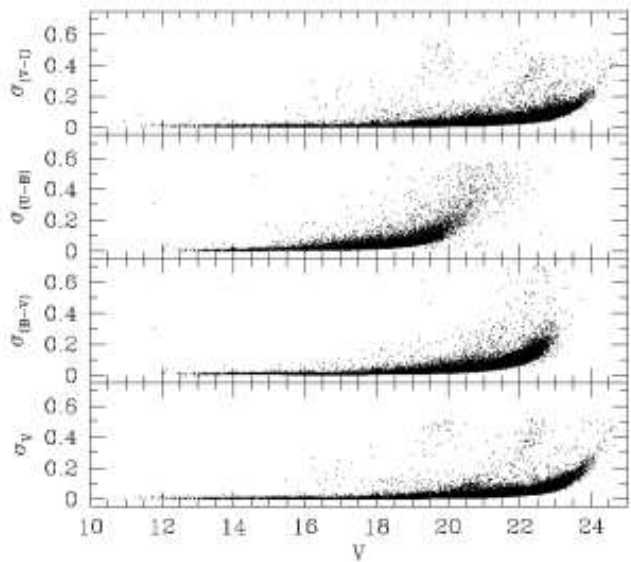}
\caption{Trend of the photometric errors in V, (V-I), (B-V) and (U-B)
as a function of the magnitude V.}
\label{ccd}
\end{figure}

\begin{figure}
\centering
\includegraphics[width=9cm]{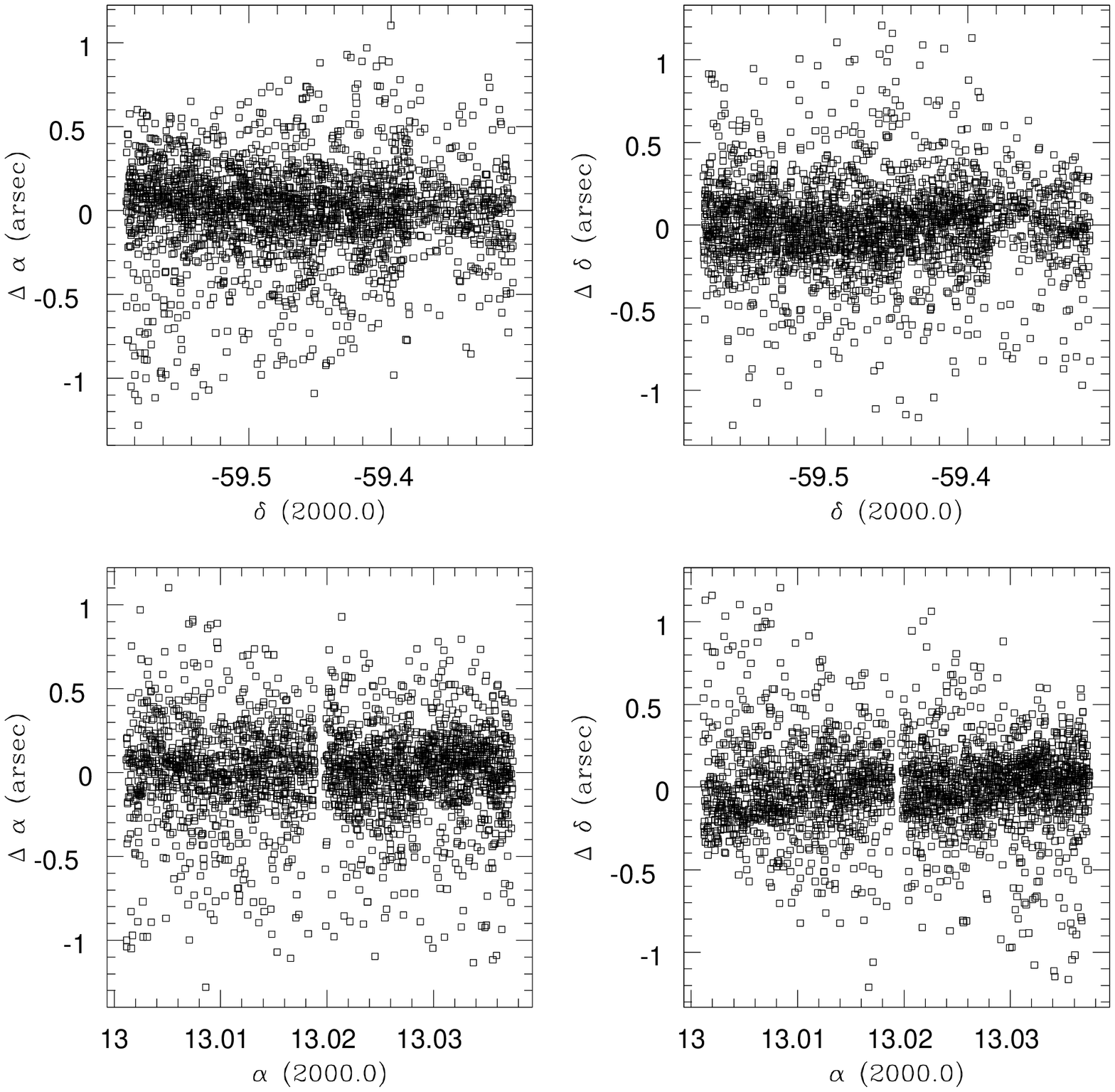}
\caption{Positional residuals on the right ascension and declination
transformations obtained from a comparison of our astrometric solution
and the GSC 2.2 reference catalog as a function of RA and Dec for
all the eight CCDs of the WFI mosaic.}
\label{ccd}
\end{figure}

\section{Introduction}
This paper is the second on a series (Prisinzano et al. 2001, hereinafter
Paper I) devoted to the study of the stellar
Luminosity (LF) and Mass Functions (MF) in Galactic Open Clusters as a function
of age and position in the Galactic disk. The final goal is to probe
the environmental dependence of the Initial Mass Function (IMF) in open star
clusters.\\
Unfortunately, this is a challenging task for several reasons.
First, most open clusters are poorly populated ensembles of stars,
containing between tens to hundreds of stars, and only in rare cases
up to some thousands.  Secondly, they are located in the inner regions
of the thin disk. This fact makes it very difficult to segregate
cluster members (at odds with globular clusters), due to the high
contamination from field stars located in the line of sight of the
clusters. Third, they can be strongly obscured due to interstellar
absorption between us and the cluster.\\
Therefore, to obtain cluster membership is a cumbersome process.
This is feasible for clusters close to the Sun, for which we can obtain
high quality radial velocity and proper motion measurements. Going
farther away one has to rely on statistical corrections which are
usually made by comparing the cluster with the Galactic disk field in its outskirts
(see Paper I, Kalirai et al. 2003 and Burke et al. 2004 for similar studies).\\
Ideally, when a correction for field star contamination has been
performed, a reasonable Present Day Mass Function (PDMF) can
be derived from the observed LF.  By modeling the dynamical evolution
of the cluster the PDMF can be converted into an IMF which can be
compared with the MFs of other clusters to look for universality or
deviations.\\
In Paper I we discussed
the intermediate-age open cluster NGC~4815
($l = 303^{\circ}.63$, $b = -2^{\circ}.09)$, 
for which we found that
the PDMF presents a slope $\alpha \approx 3$ in the mass range $2.5
\leq\frac{m}{m_{\odot}} \leq 0.8$.  
In this paper we present results
for NGC~4852 (VDBH 143, $l = 304^{\circ}.03$, $b = 3^{\circ}.25$), a
seemingly younger open cluster never studied insofar, observed with the WFI
camera attached to the 2.2m telescope in La Silla.\\
The size of the eight CCDs mosaic allows us also
to obtain a good representation of the field stellar
population in the direction of the cluster to be used
for a statistical subtraction of foreground/background objects.\\
The layout of the paper is as follows. In Section~2 we present the
observations and data reduction technique; 
Section~3  presents a detailed analysis of the star
counts, and a determination of the cluster size, while Section ~4 is dedicated to
a discussion of the CMDs.
Section~5
illustrates the derivation of the cluster's basic
parameters. Section~6 deals with the LF, while Section~7 presents the
MF of NGC~4852 and the derivation of the cluster mass. Finally, Section~8
summarizes our results.

%______________________________________________________________________________

\section{Observations and reduction\label{obs}}

\subsection{Observations \label{data}}

The open cluster NGC~4852 has been observed using the Wide Field
Imager (WFI) camera mounted at the Cassegrain focus of the MPG/ESO
2.2m Telescope at La Silla (Chile). This camera is a $4 \times 2$
mosaic of 2K $\times$ 4K CCD detectors (see Fig.~1). The scale is $0\farcs238$/pix,
therefore the mosaic covers $34^{'} \times 33^{'}$ and, due to the
narrow inter-chips gaps, the filling factor is 95.9\%. Data were
obtained in the photometric night of July 7, 2002 with seeing values
near to $1^{\prime\prime}$. Details of the observed fields and
exposure times are listed in Table~1.

\begin{table} [htb]
\tabcolsep 0.25truecm
\caption {Log-book of the observations of the open cluster NGC~4852,
July 7, 2002.
The field was centered at RA =  13:00:06, DEC =-59:36:02. }
\begin{tabular}{lrcccc} 
\hline
\hline
\multicolumn{1}{c}{Field} &
\multicolumn{1}{c}{Airmass} &
\multicolumn{1}{c}{Filter}&  
\multicolumn{1}{c}{Exp. Times [sec]} \\
\hline
NGC~4852       &  1.239 & $V$ & 1, 30,  900 \\
	       &  1.294 & $I$ & 1, 30,  600 \\
	       &  1.196 & $B$ & 1, 60, 1200 \\
	       &  1.169 & $U$ & 1, 30,  600 \\
\hline
\hline
\end{tabular}
\end{table}

\subsection{Data reduction and calibration}

All the images have been pre-processed in a standard way with the
IRAF\footnote{IRAF is distributed by NOAO, which are operated by AURA
under cooperative agreement with the NSF.} package CCDRED and using
the sets of bias and sky flat-field images collected during the same
observing night.  Instrumental magnitudes and positions of the stars
for each frame have been derived by profile-fitting photometry with
the DAOPHOT package, using the Point Spread Function (PSF) method
(Stetson 1987). In order to obtain the transformation equations
relating the instrumental ($u,b,v,i$) magnitudes to the standard
$UBV$ (Johnson), $I$(Kron-Cousins) system, we followed the procedure
already described in Baume et al. (2004).  Four Landolt (1992) fields
of standards have been observed. Specifically: SA-092 (23 stars),
SA-104 (20 stars), SA-107 (19 stars) and SA 110 (23 stars). For all of
these stars, aperture photometry was obtained on all the images. The
transformation coefficients, and the final calibrated photometry were
computed by using the package PHOTCAL. We used transformation
equations of the form:\\

\noindent
$ u = U + u_1 + u_2 * X + u_3~(U-B)$ \\
$ b = B + b_1 + b_2 * X + b_3~(B-V)$ \\
$ v = V + v_1 + v_2 * X + v_3~(V-I)$ \\
$ i = I + i_1 + i_2 * X + i_3~(V-I)$ \\

\begin{table} [htb]
\tabcolsep 0.5truecm
\caption {Coefficients of the calibration equations}
\begin{tabular}{ccc} 
\hline
$u_1 = -2.87 \pm 0.01$ & $u_2 = 0.46$ & $u_3 = -0.05 \pm 0.02$ \\
$b_1 =  0.36 \pm 0.02$ & $b_2 = 0.27$ & $b_3 = -0.26 \pm 0.03$ \\
$v_1 =  0.91 \pm 0.02$ & $v_2 = 0.12$ & $v_3 = +0.06 \pm 0.02$ \\
$i_1 =  2.00 \pm 0.03$ & $i_2 = 0.06$ & $i_3 = -0.16 \pm 0.02$ \\
\hline
\end{tabular}
\end{table}

\noindent
where the values of the coefficients are listed in Table~2.  In these
equations $ubvi$ are the aperture magnitude already normalized to 1
sec, and $X$ is the airmass. Second-order color terms were tried and
turned out to be negligible in comparison to their uncertainties. 
Unfortunately, a few bright stars were saturated in B and U.\\
In Fig.~2 we present the photometric errors trends as a function
of the magnitude V. The error in V  keep below 0.1 mag up to V=20.
The global  rms of the calibration are 0.02~mag both for 
all the filters.  Our photometry consists of 145,925
stars and will be made available electronically at CDS.

\subsection{Astrometry}

The astrometric solution is a basic step to combine all the eight chips
and to put all the stars in the same coordinate system.
In order to obtain an astrometric solution we use the SkyCat tool
and the Guide Star Catalogue v2 (GSC-2) at ESO. This way we find about
1500 stars in each chip for which we have both the celestial
coordinates on the GSC-2 and the corresponding pixel
coordinates. Then, by using the IRAF tasks CCXYMATCH, CCMAP and
CCTRAN, we find the corresponding transformations between the two
coordinate systems for each chip and compute the individual celestial
coordinates for all the detected stars. The transformations have an
rms value of $0\farcs29$, quite in agreement with other studies
(Momany et al. 2001, Prisinzano et al. 2004). The results are displayed
in Fig.~3, where all the reference stars are considered.

\subsection{Artificial star tests}

In order to obtain the LF, we estimated the completeness of our
sample. Completeness corrections have been determined by standard
artificial-star experiments on our data (see Piotto \& Zoccali 1999
and Baume et al. 2004).  
We selected only the stars having
fitting parameters $\chi \leq 1.0$ and $ABS({\it sharp}) \leq 0.2$, and 
located in the Color Magnitude Diagram (CMD) inside the strip shown in Fig.~8 (dashed lines,
see the discussion below).\\
Only the stars selected this way are going to be use in all the following
analysis.
Basically, we created 5 artificial images  by adding to the original images
artificial stars. In order to avoid the creation of overcrowding, in each
experiment we added at random position only 15$\%$ of the original number 
of stars. The artificial stars had the same color and luminosity distribution
of the original sample. The incompleteness, defined as the ratio of the found stars 
over the added artificial stars is listed in Table~3 for the $V$
magnitude. Since the stars in each bin has both $V$ and $I$ magnitudes,
we derive the incompleteness level also for the $I$ magnitude by adjusting
the $I$ bins according to the stars color.
Due to the relatively 
low crowding the magnitude migration effects (Piotto \& Zoccali 1999) were
negligible. A good rule of thumb is that LF counts should not be corrected by a
factor greater than 2.

\begin{table} [htb]
\tabcolsep 0.5truecm
\caption {Completeness analysis results}
\begin{tabular}{cc} 
\hline
\hline
$\Delta V$ & NGC~4852 \\
mag & \% \\
\hline
15-16 &   99.8$\pm$15.4   \\
16-17 &   97.1$\pm$14.0   \\
17-18 &   97.6$\pm$13.7   \\
18-19 &   97.5$\pm$ 9.6   \\
19-20 &   93.9$\pm$ 7.6   \\
20-21 &   86.5$\pm$ 5.9   \\
21-22 &   88.2$\pm$ 4.5   \\
22-23 &   88.7$\pm$16.0   \\
23-24 &   69.1$\pm$71.6   \\
\hline
\hline
\end{tabular}
\end{table}

In the LFs we include only the values for which the completeness
corrections, defined as the ratio between the number of found
artificial stars to that of the original added ones, was 50\% or
higher. Moreover we do not consider the bin between 23 and 24 mag.,
since the error is huge. Therefore we set the limiting magnitude to V = 23.

\begin{figure}
\centering
\includegraphics[width=9cm,height=14cm]{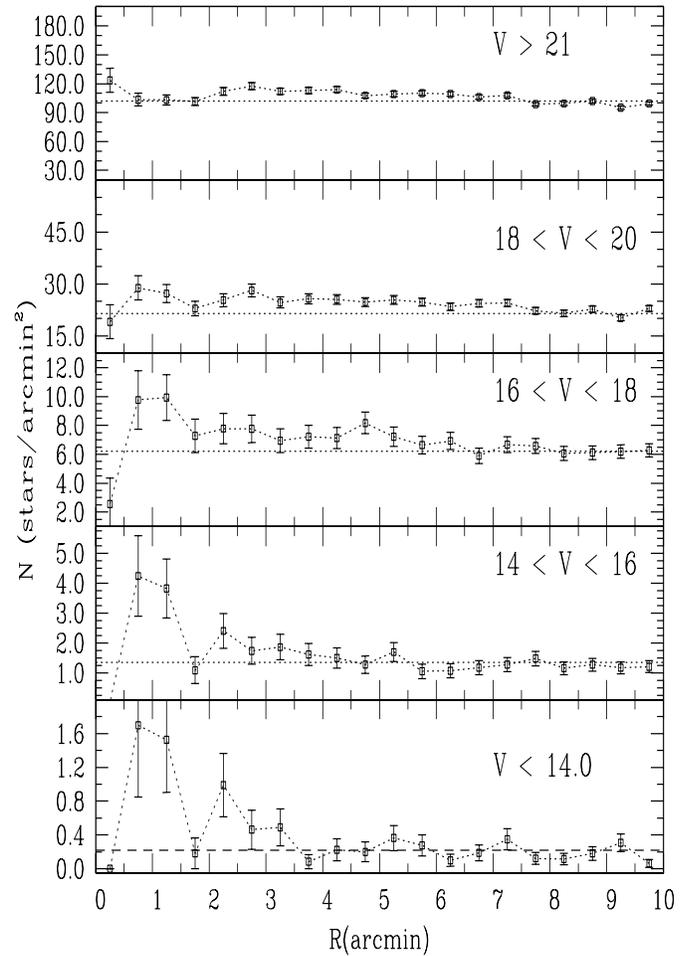}
\caption{Star counts as a function of radius from the adopted cluster
center for various magnitude intervals. The dashed line in each panel
indicates the mean density level of the surrounding Galactic disk
field in that magnitude level.}
\end{figure}

\begin{figure*}
\centering
\includegraphics[width=15cm]{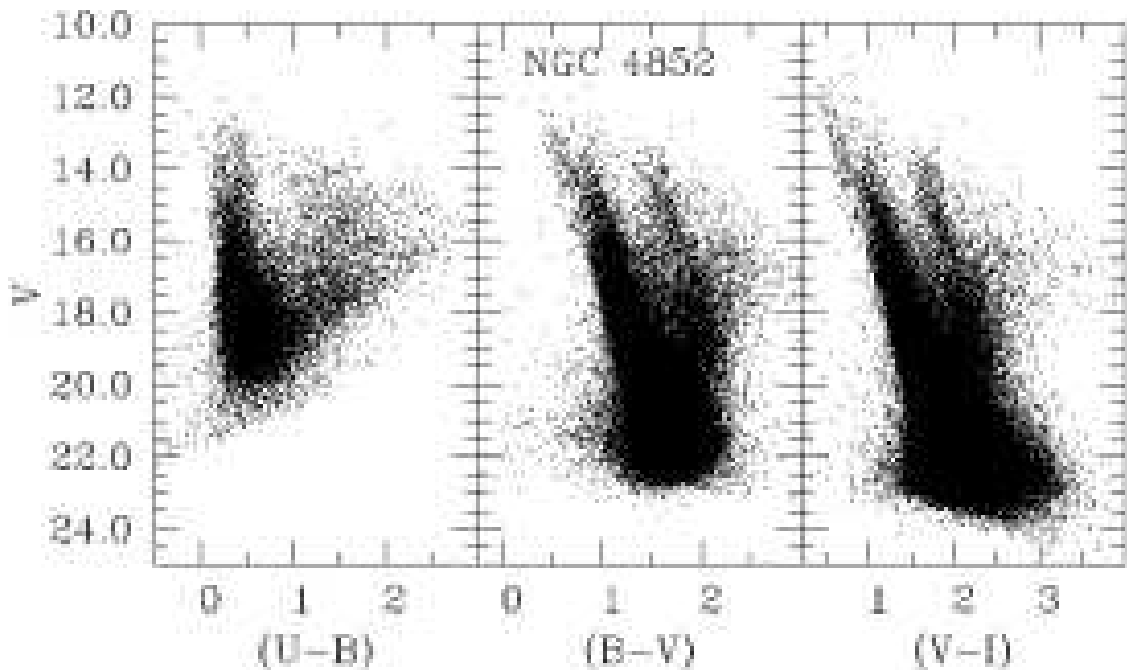}
\caption{Mosaic CMDs of NGC~4852.}
\end{figure*}

\section{Star counts and cluster size}
The aim of this section is to obtain the surface density distribution
of NGC 4852, and derive the cluster size in the
magnitude space by means of star counts. The cluster radius is indeed
one of the most important cluster parameters, useful (together with
cluster mass) for a determination of cluster dynamical parameters.
Star counts allow us to determine statistical properties of clusters (as
visible star condensations) with respect to the surrounding stellar
background.\\
By inspecting Fig.~1, NGC~4852 appears as a concentration of bright stars
in a region of about 4-5 arcmin.
In order to derive the radial stellar surface density we first seek for the
highest peak in the stellar density to find out the cluster center.
The adopted center is placed at $\alpha = 13:00:06.5$; $\delta =
-59:36:02.0$, which is very close to that given by Dias et
al. (2002). Then, the radial density profile is constructed by
performing star counts inside increasing concentric annuli $0\farcm5$
wide, around the cluster center and then by dividing by their
respective surfaces. This is done as a function of apparent magnitude,
and compared with the mean density of the surrounding Galactic field
in the same brightness interval. The contribution of the field
has been estimated through star counts in the region outside 9 arcmin
from the cluster center. Poisson standard deviations have also been
computed and normalized to the area of each ring as a function of the magnitude,
both for the cluster and for the field.\\
The result is shown in Fig.~4, where
one readily sees that NGC~4852 significantly emerges from the mean
field above V$\approx$20. At fainter magnitudes the cluster is 
confused with the Galactic disk population. Based on the radial density
profiles in Fig.~4, we find that stars brighter than V=16 
provide a cluster radius smaller than 3.0 arcmin, whereas 
in the interval $16 \leq V \leq 20$ the cluster radius
is somewhat larger (around 5.5 arcmin). 
This situation is  compatible with the cluster having experienced 
some mass segregation,
as a consequence of which massive stars sinked toward the cluster
center, while lower mass stars got spread toward the cluster envelope.
Therefore we propose that the cluster possesses a core-corona structure
conseguence of dynamical evolution, with most of the bright stars
located inside the core (3 arcmin in radius).\\

\noindent
We adopt as a final estimate
of the radius the values $5.0\pm1.0$ arcmin. This is 
larger than the estimate reported by Dias et al. (2002),
which was simply based on a visual inspection.  We shall adopt this values of
the cluster radius throughout this work. \\
We stress however that this
radius is not the limiting radius of the cluster, but the distance
form the cluster center at which  the cluster population starts to be confused
with the field population.

\section{The Color-Magnitude Diagrams\label{cmd}}

We obtained UBVI photometry with the aim of determining the LF
(and MF) of the Main Sequence (MS) stars in NGC~4852. We used four colors
as we wanted also to derive Two Color Diagrams  
and the CMDs, which allow us to: {\it i)}
discriminate stars from false detections, {\it ii)} better
identify the cluster population, in particular the MS stars, 
and {\it iii)} derive  estimates of the cluster fundamental
parameters.\\ 
The mosaic CMDs for NGC~4852 are shown in Fig.~5 for the V vs (U-B),
V vs (B-V) and V vs (V-I) combination, in the left, middle and right panel,
respectively. 
Our photometry reaches V = 23.5 at the base of the MS in the
V vs (V-I) diagram (right panel).\\ The MS
extends almost vertically for more about 10 magnitudes, from V = 13 to
V=23.5, although the completeness analysis prevents us
from using stars dimmer than V= 23.  The width of the MS
gets larger at fainter magnitudes. This is partly due to the
increasing photometric errors at increasing magnitudes (see Fig.~2).  
However the MS is much wider than expected simply from photometric errors.\\
Between the various causes which concur to enlarging the natural MS
width we can envisage the presence of unresolved binary stars (common
in open clusters), the contamination by foreground stars, 
a possible spread in metallicity, and the 
differential reddening across the cluster area. To have an indication
of the importance of differential reddening we used FIRB extinction
maps from Schlegel et al. (1998). Across the 30 squared arcmin 
covered by our photometry the typical reddening variations amount to
0.2 mag. (E(B-V) in the range 0.70-0.90 mag, but see also next Sect.).\\
Therefore we conclude that differential reddening is not the major
cause of the MS broadening, but this is probably mainly due to the
contamination from foreground field stars.\\
The CMDs in Fig.~5 show a red sequence which detaches from the MS
at V $\approx$ 19 (middle and right panels)  and which corresponds to
the Red Giant Branch of the Galactic disk
population. 
Besides, at about V= 14.5 there seems to be a bifurcation which deserves
further attention.\\
\noindent

\begin{figure*}
\centering
\includegraphics[width=14cm,height=12cm]{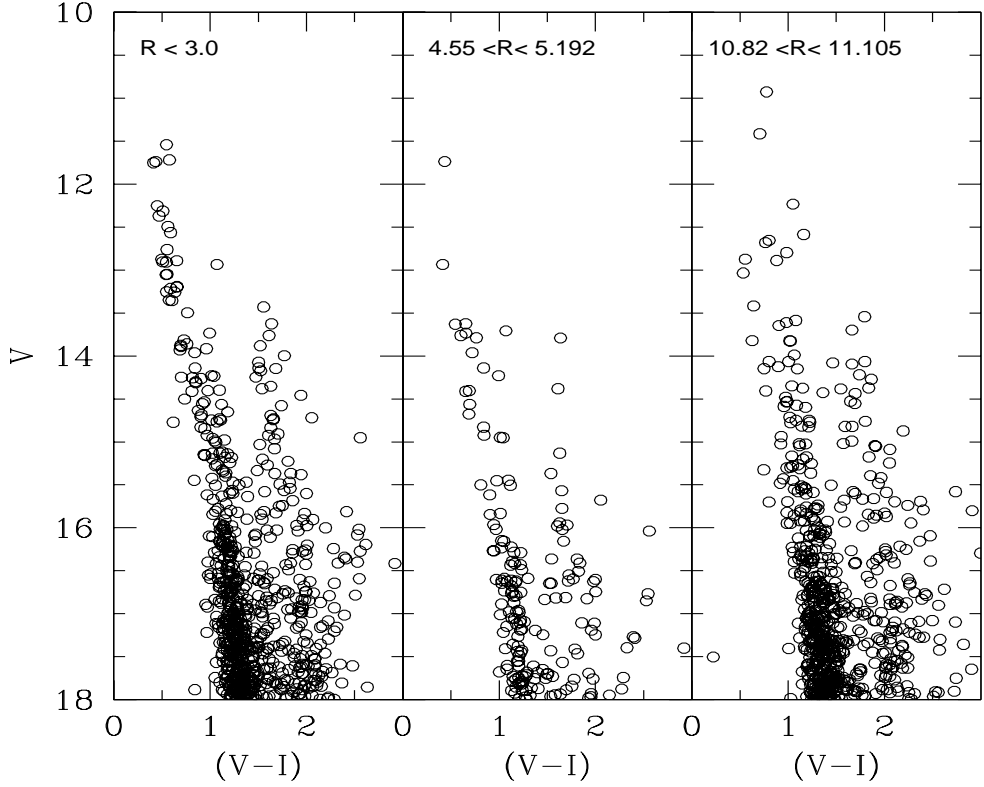}
\caption{A zoom of the upper part of the $VI$ CMD for NGC 4852 
in different spatial location. {\bf Left panel:} Stars with 3.0 arcmin.
{\bf Middle panel:} Stars at $4.55 \leq r[arcmin] \leq 5.192$.
{\bf Right panel:} Stars at $10.82 \leq r[arcmin] \leq 11.105$.}
\end{figure*}

\noindent
To better clarify this issue, we provide in Fig.~6 a zoom of the 
upper MS of the CMDs in the right panel of Fig.~5 (V vs (V-I)). 
In the left panel we show the MS of
all the stars located within  3.0 arcmin from the nominal cluster center.
In the middle panel we show the stars located at $4.55 \leq r[arcmin] \leq 5.192$,
whereas in the right panel we show the CMD of the field population, sampled
outside the cluster radius, at $10.82 \leq r[arcmin] \leq 11.105$. The limits
have been selected in order to keep fixed the sampling area.\\
\noindent
Fig.~6 confirms the findings of star counts (see previous Sect.).
Most of the  cluster bright stars (V $\leq$ 14) population is located 
inside 3.0 arcmin (left panel). This population does not
have a counterpart in the field population presented in the right panel.
The few bright stars in this CMD in fact are much redder than the 
mean MS color at the same mag. level in the left panel CMD.\\
However a few bright MS stars are still present outside the cluster inner region
(middle panel). \\
\noindent
On the other hand, stars fainter than V $\approx$ 16 are present also
outside the cluster central region (middle and right panels), 
as already emerged from star counts.
The group of star in the right panel at 15 $\leq V \leq$ 15.5
appear as well redder than the mean MS color at this magnitude
and we are keen to believe that they are field stars. \\
\noindent
The confusion with the stellar field is so high that we cannot conclude
anything about the cluster population below V $\approx$ 17.\\
The overall morphology of these CMDs confirm the mass segregation scenario
emerged from star counts
and provides an estimate of the radius around 5 arcmin.\\

\noindent
In summary, Fig.~6 suggests that
NGC~4852 is indeed an open cluster, presumably young,
highly contaminated, which underwent dynamical evolution and mass
segregation.\\
Finally, the bifurcation visible in Fig.~5 can be explained as the contribution
of the cluster population, responsible for the blue arm, 
and the contamination of foreground
stars, responsible for the red arm.

\begin{figure}
\centering
\includegraphics[width=9cm]{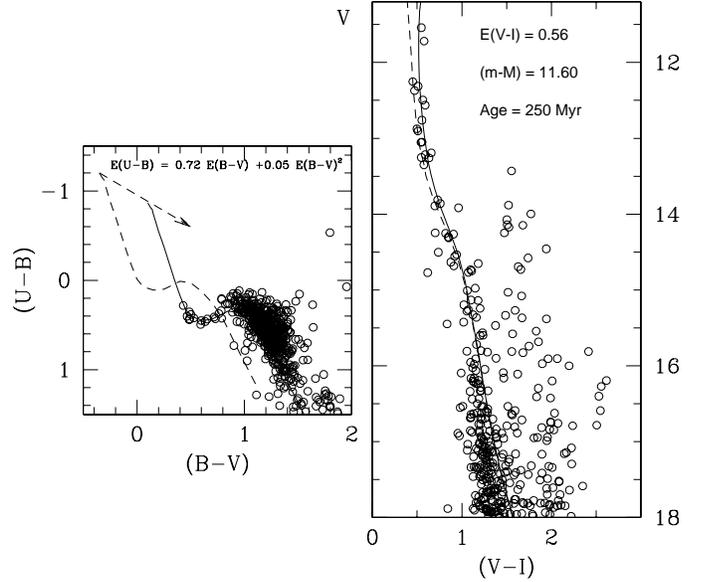}
\caption{{\bf Left panel:} Two color diagram for the
NGC~4852 (star within the cluster radius). The dashed line is an empirical Schmidt-Kaler (1982)
ZAMS, while the solid line is the same ZAMS shifted by
E(B-V) = 0.45, E(U-B) = 0.72 $\times$ E(B-V) + 0.05 $\times$
E(B-V)$^2$. The dashed arrow indicates the reddening vector.
{\bf Right panel:}
The CMD of NGC 4852 (star within the cluster radius). The dashed line
is an empirical Schmidt-Kaler (1982)
ZAMS shifted by E(V-I) = 0.56 and (m-M) = 11.60, while the solid line is a solar
metallicity isochrone for the age of 200 Myr.}
\end{figure}

\section{Cluster fundamental parameters}
In order to estimate the cluster's fundamental parameters, we
consider only the stars located inside 5.0 arcmin
and fulfilling the selection criteria defined
in Sect.~2. This is because we want to limit the star contamination
from the field and because the cluster MS is better defined
by the brightest stars.\\
These stars are shown in the Fig.~7. 
In the left panel we present the fit in the Two Color Diagram, which provides
a reddening E(B-V) =0.45$\pm$0.05 (error by eye).
Unfortunately, a few bright stars were saturated in B and U,
and therefore the sequence is populated only starting from
a spectral type of about B8. Nonetheless, the fit
is very good, and allows us also to see that the
amount of differential reddening is small, of the order
of 0.1 mag, as suggested also by FIRB maps (see Sect~4).\\
We derived also E(V-I) by using the ratio
$\frac{E(V-I)}{E(B-V)}=1.244$ from Dean et al. (1978),
and it results to be $0.56\pm0.05$ mag., and it turns
out to perfectly match the ZAMS fitting estimate in the right panel
of the same figure.
There we show the upper MS in the V vs (V-I) plane.
Over-imposed 
is an empirical ZAMS (dashed line) from
Schmidt-Kaler (1982, dashed line)) 
shifted by (m-M)$_V$ = $11.60\pm0.30$ mag., E(V-I)= $0.56\pm0.05$
mag. (errors by eye).
The empirical ZAMS well reproduces  the distribution of the stars,
especially the bulk of the bright stars, which better identify
the cluster.\\
By adopting the normal value for the total to selective absorption
(R = $\frac{A_V}{E(B-V)}$ = 3.1), we derive an absolute distance modulus
$(m-M)_0$ = 10.20, which places the cluster 1.1 kpc from the Sun.
Finally, (m-M)$_I$ turns out to be 11.0 mag.\\
\noindent
As for the age, the cluster appears 
young, with all the stars still in the MS. Only the two brightest stars in
the CMD, although at different level, seem to be in the act of
leaving the MS. We have tried to over-impose a few  
solar metallicity isochrones from
the Padova models (Girardi et al. 2000), for ages ranging from 10 to 300
Myr, and found that 
the 200 Myr isochrone (solid line in Fig.~7, right panel) 
accounts very well for the brightest star.
Therefore we are keen to believe that the cluster cannot be older 
than this age.

\begin{figure}
\centering
\includegraphics[width=9cm]{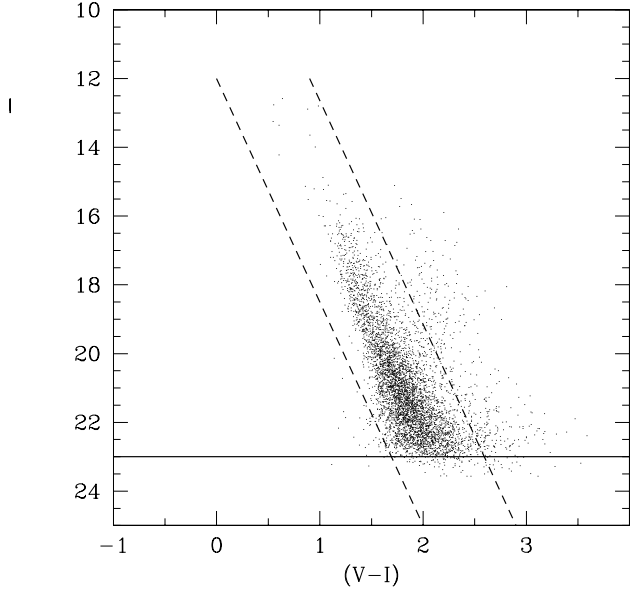}
\caption{The CMD of NGC 4852 stars within the 
adopted cluster radius (5 arcmin). The solid horizontal line
indicates the completeness limit, whereas the two dashed
tilted lines confine the region where we consider
MS stars in the cluster are more probably located.}
\end{figure}

\section{Luminosity Function}

The completeness analysis performed in Section~2 allows us to derive the
LF.
This has been constructed as follows:\\

\noindent
({\it i}) We selected stars within the adopted cluster radius (see Sect~3) and 
according to the fitting parameters (see Sect.~2). Moreover
we used only the MS stars, which
we looked for by considering the stars located within the two ridge-lines
shown in Fig.~8, defined by the two relations
V=6.5(V-I) + 17.8 and V=6.5(V-I) + 12.0. 
This is mainly 
to avoid contamination coming from the
Galactic disk RGB population (see also Fig.~5). \\

\noindent
({\it ii}) We used the standard histogram
technique, and adopted bins 0.5 mag wide. 
Therefore we counted the number of stars in the cluster
area starting from V = 12.0 mag. The contribution
from the Galactic field was derived
outside the cluster region
from a corona  with the same area of the cluster.
Also for the field we counted the number of stars in the corona
starting from V = 12.0 mag.\\
After having completeness
corrected (see Table~3) both the cluster and the field counts, 
we derived the NGC 4852 LF by subtracting, bin by  bin,
the field area counts from the cluster area ones.
At this point we computed the logarithm of the cumulative
counts distribution, say the logarithm  of the field-corrected
number of stars fainter than a given magnitude.\\

\noindent
These are presented  in Fig.~9 
where we plot
the logarithm of the cumulative
counts distribution (log Ncum) as a function of the magnitude V
for the total sample (dashed line), field stars (dotted line)
and the LF (filled triangles).\\
The solid line passing through the filled triangles is the resulting LF,
and the two other solid lines below and above are the lower an upper boundaries
of the LF due to errors in the LF derivation.\\
These errors take into account the Poisson errors of the
counts both in the cluster and in the field population,  
and the errors derived from the completeness corrections.  
We adopted (see Table~3) a limiting magnitude V = 23 
as completeness limits.  
\noindent
The resulting LF is a raising function
down to V $\approx$ 20 mag. Downward the contributions of the stars
from the cluster and the field are basically the same, and the LF does not 
increase significantly anymore.

\begin{figure}
\centering
\includegraphics[width=9cm]{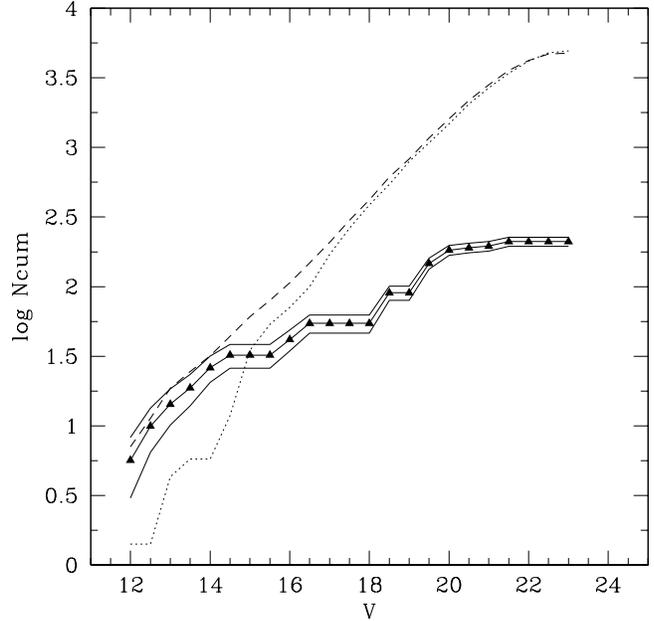}
\caption{Luminosity Function (solid triangles) 
in the V band of NGC 4852 after incompleteness
correction and statistical field stars subtraction. The contribution
of the total sample and the field stars are shown with a  dashed and a dotted
line, respectively.}
\end{figure}

\section{Mass Function}

The LFs can be transformed into MFs using a mass-luminosity relation
(MLR).  Since we could not obtain an empirical transformation, we must
rely on theoretical models. Therefore, we used the ZAMS relation
by Girardi et al. (2000) with a metallicity Z=0.020. The distance
modulus and the reddening are those derived in Sect.~5\\
\noindent
The resulting MF, obtained from both the LF, is shown in
Fig.~10. 
The MF is monotonically increasing up to  $log(\frac{m}{m_{\odot}})=-0.2$ (say
$M = 0.60 M_{\odot}$).
This  corresponds to V = 20.0, which by the way is roughly
the magnitude at which the cluster melts with the field
(see Fig.4). Afterward, the MF does not increase significantly anymore.\\
\noindent
Over the mass range $3.2 \leq\frac{m}{m_{\odot}} \leq 0.6$  
($12 \leq V \leq 20$)
the MF is a clear power law with a slope $\alpha = 2.3\pm0.3$,
basically consistent with the Salpeter one,
which in our notation has a slope of 2.35.\\

\noindent
Finally, it is possible to obtain an estimate of the cluster mass by
integrating the luminosity function. In detail,

\begin{equation}
\label{mass}
M=<m>\times N=N\int \limits_{M_{V_{min}}}^{M_{V_{max}}} m(M_V)\phi (M_V)dM_V
\end{equation}

where $M_V$ is absolute stellar magnitude $M_V=V-(m-M)_V$, $\phi(M_V)$
is the normalized luminosity function

\begin{equation}
\label{phi}
\phi(M_V)=f(M_V)/N ,
\end{equation}

\noindent
N is the total number of stars, $m(M_V)$ is the adopted mass-luminosity
relation. The luminosity function and mass-luminosity relation have
been approximated by a spline.
The limits in the integral are $M_{V_{min}}=0.4$ (V = 12) 
and $M_{V_{max}}=8.4$ (V=20). The integral turns out to be:

$$M\simeq 2570\pm 210 M_{\odot}$$

\noindent
This mass estimate refers to the limiting magnitude of $V=20.0$,
and represents a lower limit of
the cluster total mass, due to the effects of the field stars
contamination.

\noindent
The error is determined from mass estimate for upper and lower boundaries
of the LF (see Fig.9). 

\begin{figure}
\centering
\includegraphics[width=9cm]{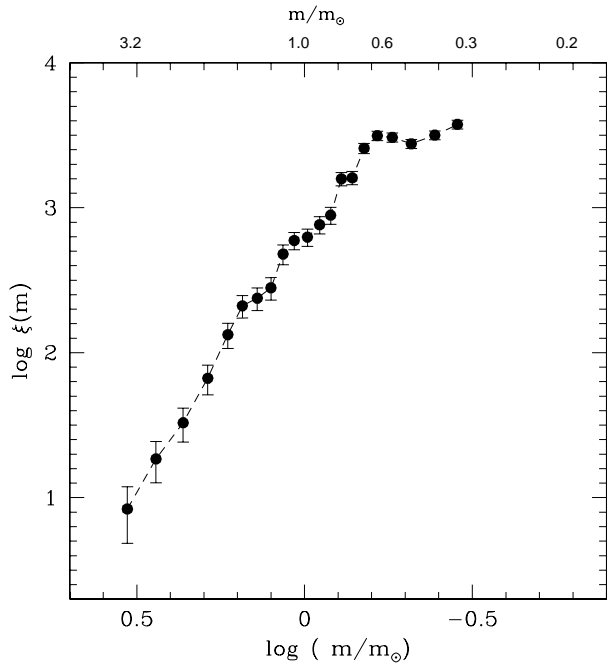}
\caption{The MFs of NGC 4852 derived from LFs}
\end{figure}

\section{Discussion and conclusions}
In this paper we have presented the first study of the young open
cluster NGC~4852, for which we have obtained wide field deep $UBVI$
CCD photometry.
The cluster turns out to be young, with a maximum
age of 200 Myr, and it is located at 1.1 kpc from the Sun.\\
We have derived LFs and present day MF for the cluster.
The MF is one of the most extended in mass insofar
obtained (see Paper I, Fig.~9),  and we find
that the MF slope is basically consistent  with the classical Salpeter
one over the whole mass excursion, without any significant slope
change.
According to Scalo (1998) the mean slope for the MF in
clusters younger than half a Gyrs is $\alpha = 2.25\pm0.88$,
and NGC~4852 reflects this behavior.
Below 0.6$m_{\odot}$, the MF is not reliable anymore,
since the cluster is completely confused with the general
Galactic disk field.

%%%%%%%%%%%%%%%%%%%%%%%%%%%%%%%%%%%%%%%%%%%%%%%%%%%%%%%%
\begin{acknowledgements}
G. Carraro deeply thanks Y. Momany and E. Held for many fruitful
discussions and acknowledges financial support from ESO during a visit
to Vitacura (Santiago) and from the {\it Fundaci\'on Andes}. 
R. A. M\'endez acknowledges support from the
Chilean {\sl Centro de Astrof\'\i sica} FONDAP No. 15010003.
The work of G. Baume has been supported by Padova University
through a postdoctoral grant.
\end{acknowledgements}
%%%%%%%%%%%%%

%%%%%%%%%%%%%

\begin{thebibliography}{}
\bibitem{}Baume G., Moitinho A., Giorgi E.E. et al. 2004, 
          A\&A 417, 961
\bibitem{}Baume G., V\'azquez R.A., Carraro G. et al. 2003, 
          A\&A 402, 549
\bibitem{}Burke, C.J., Gaudi B.S., DePoy D.L. 2004, AJ 127, 2382
\bibitem{}Dean J.F., Warren P.R. \& Cousins A.W.J. 1978, MNRAS 183, 569
\bibitem{}Dias W.S., Alessi B.S., Moitinho A., et al 2002, A\&A 389, 871
\bibitem{}Girardi L., Bressan A., Bertelli G., \& Chiosi, C. 2000, A\&AS 141, 
          371
\bibitem{}Kalirai J.S., Fahlmam G.G., Richer H.B. et al.,
          2003, AJ 126, 1402
\bibitem{}Landolt A.U., 1992, AJ 104, 340
\bibitem{}Momany Y., Vandame B., Zaggia S. et al., 2001, 
          A\&A 379, 452
\bibitem{}Piotto G. \& Zoccali  M. 1999, A\&A 345,485
\bibitem{}Prisinzano L., Carraro G., Piotto G. et al. 2001, A\&A 369, 851
\bibitem{}Prisinzano L., Micela G., Sciortino S. et al., 2004
          A\&A 417, 945
\bibitem{}Salpeter E.E., 1955, ApJ 129, 608
\bibitem{}Scalo J.J. 1998, in ASP .Conf. Ser. 142, The Stellar Initial
          Mass Function, 38th Herstmonceux Conference, 201+
\bibitem{}Schlegel D.J., Finkbeiner D.P, Davis M. 1998, ApJ 500, 525   
\bibitem{}Schmidt-Kaler Th. 1982, Landolt-B\"ornstein,
          Numerical data and Functional Relationships in Science and
          Technology, New Series, Group VI, Vol. 2(b), K. Schaifers
          and H.H. Voigt Eds., Springer Verlag, Berlin, p.14
\bibitem{}Stetson P.B. 1987, PASP 99, 191
\end{thebibliography}
\end{document}